\documentclass[12pt]{article} 

\usepackage{url}
\usepackage{setspace}
\usepackage{amsmath,amssymb,amscd}
\usepackage{amsfonts}
\usepackage{color}
\usepackage{graphicx}
\usepackage{braket}
\usepackage{commath}
 \usepackage{cite}

 \newcommand{\bea}{\begin{eqnarray}}
\newcommand{\eea}{\end{eqnarray}}
\newcommand{\be}{\begin{equation}}
\newcommand{\ee}{\end{equation}}
\newcommand{\ba}{\begin{align}}
\newcommand{\ea}{\end{align}}

\newcommand\rref[1]{(\ref{#1})}

\newcommand{\zb}{\bar{z}}
\newcommand{\hb}{\bar{h}}

\newcommand{\ie}{{\it i.e.~}}

\newlength{\slength}
\newcommand*{\bb}{\makebox[\slength][c]{$\bullet$}}
\newcommand*{\wb}{\makebox[\slength][c]{$\circ$}}

\begin{document}

\begin{titlepage}
\begin{center}

\hfill \\
\hfill \\
\vskip 0.75in

{\Large \bf  Permutation Orbifolds and Chaos}\\

\vskip 0.4in

{\large Alexandre Belin } \\
\vskip 0.3in

{\it Institute for Theoretical Physics, University of Amsterdam,
Science Park 904, \\ 1098 XH Amsterdam, The Netherlands} \vskip 3.5mm

\texttt{a.m.f.belin@uva.nl}

\end{center}

\vskip 0.35in

\begin{center} {\bf ABSTRACT } \end{center}
We study out-of-time-ordered correlation functions in permutation orbifolds at large central charge. We show that they do not decay at late times for arbitrary choices of low-dimension operators, indicating that permutation orbifolds are non-chaotic theories. This is in agreement with the fact they are free discrete gauge theories and should be integrable rather than chaotic. We comment on the early-time behaviour of the correlators as well as the deformation to strong coupling.

\vfill

\noindent \today

\end{titlepage}

\tableofcontents

\section{Introduction and Summary}
Quantum chaos has recently received a surge of interest in the context of holography, which has led to new insights on thermal physics of quantum gravity and conformal field theories. This program was initiated by a holographic realization of the butterfly effect \cite{Shenker:2013pqa}, where it was shown that a small boundary perturbation can have drastic consequences in the bulk provided it happens at sufficiently early times. This effect along with several generalization were then studied in various quantum systems relevant for gravitational physics \cite{Shenker:2013yza,Leichenauer:2014nxa,Kitaevtalks,Roberts:2014isa,Roberts:2014ifa,Jackson:2014nla,Shenker:2014cwa,Maldacena:2015waa,Polchinski:2015cea,Caputa:2015waa,Hosur:2015ylk,Stanford:2015owe,Gur-Ari:2015rcq,Berkowitz:2016znt,Polchinski:2016xgd,Fitzpatrick:2016thx,Michel:2016kwn,Perlmutter:2016pkf,Turiaci:2016cvo,Caputa:2017rkm}.

In classical physics, the Poisson bracket $i\hbar\{q(t),p(0)\}$ measures the sensitivity of $q(t)$ to the initial conditions and hence diagnoses classical chaos. In analogy, one can consider the commutator
\be
\left[W(t),V(0)\right] \,,
\ee
which measures the perturbation by $V$ on measurements of $W$. If the time separation is small, this commutator will be small. However, it can grow quickly if the quantum system is chaotic. In this paper, we will consider a quantity closely connected to this commutator \cite{Maldacena:2015waa}, the out-of-time-ordered (OTO) correlation function
\be \label{OTO0}
F(t)=\frac{\Braket{V(0)W(t)V(0)W(t)}_\beta}{\Braket{VV}_\beta\Braket{WW}_\beta} \,.
\ee
The behaviour of this Lorentzian correlation function was argued to be a sharp diagnostic of quantum chaos \cite{Maldacena:2015waa}. In particular, it should decay at late times in chaotic conformal field theories for arbitrary choice of "simple" operators $V$ and $W$. Simple means that $V$ and $W$ should be a product of an $\mathcal{O}(1)$ number of degrees of freedom.

For large $N$ theories, one can distinguish two parametrically distinct time-scales. First, there is the dissipation time $t_d\sim\beta$ which is the characteristic time scale of the exponential decay of two point functions $\Braket{V(0)V(t)}_\beta$ in the thermal state. This will also be the time scale at which typical \textit{time-ordered} correlation functions reach there late time limits, for example
\be
\braket{V(0)V(0)W(t)W(t)}\sim \Braket{VV}\Braket{WW} \,, \quad \text{for} \ t> t_d \,.
\ee
For holographic theories, this time scale will be connected to the behaviour of the quasinormal modes of black holes. Next, there is the time-scale at which the out-of-time-ordered correlation function becomes small in chaotic theories, which is named the scrambling time $t_s$. For large $N$ theories, the scrambling time is parametrically larger than the dissipation time
\be
t_s\sim \log N \gg t_d \,.
\ee

A way to characterize the strength of chaos is by looking at the behaviour of $F(t)$ for times between $t_d$ and $t_s$. Chaotic theories are expected to exhibit an exponential growth in this regime which takes the form
\be \label{earlytimes}
\Braket{V(0)W(t)V(0)W(t)}_\beta\sim \left(1- \frac{1}{N} e^{\lambda_L t}\right) \,,
\ee
where $\lambda_L$ is called the Lyapunov exponent. The first term in the r.h.s. of \rref{earlytimes} corresponds to the disconnected part of the four-point function, whereas the second term is the connected contribution. We see that the connected contribution starts out to be small but becomes of the same order as the disconnected part for $t\sim t_s$. The $1/N$ hierarchy between connected and disconnected contributions is a result of large $N$ factorization and always remains valid for Euclidean correlators. Chaos can also be viewed as a breakdown of large $N$ factorization for large Lorentzian times.

In \cite{Maldacena:2015waa}, a bound on the Lyapunov exponent was found to be
\be
\lambda_L \leq \frac{2\pi}{\beta} \,,
\ee
which is saturated by black holes in Einstein gravity, showing support for the claim that black holes are the fastest scramblers in nature \cite{Sekino:2008he,Lashkari:2011yi}. This bound on chaos can also be used to carve out from the space of all CFTs, those that have nice enough properties to have potential Einstein gravity duals\cite{Perlmutter:2016pkf,Afkhami-Jeddi:2016ntf}. We will be interested in the case of two-dimensional CFTs where it was shown that identity block domination of the correlation function \rref{OTO0} yields maximal chaos \cite{Roberts:2014ifa}.

In this paper, we will study out-of-time-ordered correlation function in a large class of theories: permutation orbifolds. Starting from a two-dimensional CFT $\mathcal{C}$ with central charge $c$, we can define
\be
\mathcal{C}_N = \frac{\mathcal{C}^{\otimes N}}{G_N} \,,
\ee
for $G_N \subseteq S_N$, giving an orbifold CFT with central charge $Nc$. Provided that the group $G_N$ is oligomorphic, this will provide a vast landscape of two-dimensional CFTs that have a good large $N$ limit \cite{Belin:2014fna,Haehl:2014yla,Belin:2015hwa}. Permutation orbifolds should be viewed as free discrete gauge theories and hence provide examples of weakly-coupled CFTs. In this paper, we will show that these theories are not chaotic, in agreement with the fact that they are free gauge theories. We will show this explicitly by considering the OTO four-point function of arbitrary low-dimension operators.

Although one could have suspected this outcome on the grounds that orbifold theories are free gauge theories, there are other perspectives from which the result can appear more surprising. First, the result is completely independent of the choice of seed theory $\mathcal{C}$. This means we can pick a seed theory that is chaotic with a spectrum of operators that is, at least in principle, as crazy as we want. We can certainly make most of the operator dimensions irrational. One might suspect that this leaves an important imprint in the orbifold theory. However, our results indicate that it does not. Because the $N$ copies are non-interacting, the details of the seed theory's spectrum are completely washed out in the large $N$ limit. The details of the seed theory will only be important at early times $t\sim\mathcal{O}(1)$ but do not matter at times of order the scrambling time $t\sim\log N$.

The other reason why one could have suspected the outcome to be different is that permutation orbifolds have been capable of reproducing many universal features of Einstein gravity in the semi-classical limit. This is often tied to the dominance of the identity block in CFT correlation functions. At least for the symmetric group, the orbifolds theories give the same partition function as Einstein gravity \cite{Keller:2011xi,Hartman:2014oaa}, which also means they correctly reproduce the BTZ black hole entropy. Furthermore, one can check that they satisfy all conditions demanded in \cite{Kraus:2017kyl}, which means that finite temperature two-point functions match those calculated in the BTZ background. Finally, other observables involving late-time dynamics such as two point functions in excited states also show qualitative similarities to a theory dual to Einstein gravity \cite{Balasubramanian:2016ids} (even though there are quantitative differences in terms of time-scales). In some sense, these theories are doing a way better job at mimicking general relativity than what they should be doing. This is probably related to the fact that gravity in three dimensions is very special, while the story is more complicated in higher dimensions \cite{Belin:2016yll}. Nevertheless, these facts suggest that several observables of permutation orbifolds might be well approximated by the identity block, and pinning down which ones are and which ones aren't appears to be quite important. In this paper, we will show that the OTO correlator does not fall in this class as it behaves drastically differently from how a theory dual to Einstein gravity would behave.

In some cases (like for example when $G_N=S_N$ and $\mathcal{C}$ is the non-linear sigma model on $\mathbb{T}^4$), one can deform the orbifold theory to go to strong coupling. In that case, one expects maximal chaos which means that the Lyapunov exponent should increase as we increase the coupling. We do not perform this calculation but make some general comments on the deformation to strong coupling.

\subsection{Summary of Results}
We will show that the OTO four-point function in permutation orbifolds takes the general form
\be \label{generalform}
F(t)=1+\sum g_2(\vec{K},\vec{\varphi}) f_2(t)+\frac{1}{N}\left[\sum g_3(\vec{K},\vec{\varphi})f_3(t)+\sum g_4(\vec{K},\vec{\varphi})f_4(t)\right]+...
\ee
where the $g_i(\vec{K},\vec{\varphi})$ are order one combinatorial factors that depend on the choice of group and the choice of operators for $V$ and $W$ while the $f_i(t)$ are related to $i$-point functions of the seed theory. The sums are taken over the different $i$-point functions of the seed theory that can appear. This results follows directly from large $N$ factorization.

For times much greater than the dissipation time, we will show that the OTO correlator behaves as
\be
F(t) \approx 1+\frac{1}{N}\sum g_4(\vec{K},\vec{\varphi})f_4(t)+... \,, \quad t \gg t_d \,,
\ee
namely that the two and three-point function contributions become small and we are simply left with the seed theory four-point functions. If we consider twisted sector operators, $f_4(t)$ is not directly related to seed theory four-point functions on the plane but rather to some higher genus amplitude. Nevertheless, we can also think of its contribution as coming from four-point functions in a $S_k$ orbifold theory with $k\sim\mathcal{O}(1)$.

This will be enough to show that permutation orbifolds cannot be chaotic. There are essentially two scenarios. If the seed theory (or Sym$_k(\mathcal{C})$ for the twisted sectors) is chaotic, then the four-point function of the seed theory vanishes at lates times. In that case, $F(t)\sim 1$ at late times. If the seed theory is not chaotic, its OTO four-point function will stabilizes at some value $\alpha\sim \mathcal{O}(1)$ which means $F(t)\sim 1+ \alpha/N\sim 1$. Again, $F(t)$ does not decay. This result is universal and holds for all oligomorphic permutation orbifolds. Even if the seed theory is chaotic, the effect is washed away once we take $N$ non-interacting copies and orbifold. The orbifolding procedure does not introduce any interaction between the $N$ copies but only projects to  $G_N$ singlets and hence considerably reduces the number of low-energy states. This result should be a taken as a general feature of free gauges theories in the large $N$ limit.

The paper is organized as follows: in section 2, we discuss the kinematics of the OTO correlation functions and define notation. In section 3, we introduce permutation orbifolds and show how large $N$ factorization arises. In section 4, we study the OTO correlators in permutation orbifolds, considering both untwisted and twisted sector operators. We also discuss the behaviour of the function at early times as well as a possible deformation of the theory to strong coupling.

\section{Out-of-time-ordered Correlators in 2d CFTs}
We are interested in calculating out-of-time-ordered correlation functions in two dimensional conformal field theories. In this paper, we will focus on four point functions and consider correlation functions in the thermal state on the infinite line. We will closely follow \cite{Roberts:2014ifa} to set up our convention. 2d CFTs have the nice property that a correlation function in the thermal state on the infinite line can be mapped to usual vacuum expectation values through the map
\be \label{conftranseucl}
z(x,t_E)=e^{\frac{2\pi}{\beta}(x+it_E)} \,, \qquad \bar{z}(x,t_E)=e^{\frac{2\pi}{\beta}(x-it_E)} \,,
\ee
where $x,t_E$ label points along the spatial and thermal direction respectively. With this transformation one can easily compute Euclidean correlators by mapping the operators to the plane. The operators transform as
\be
\mathcal{O}(x,t) =\left( \frac{2\pi z}{\beta}\right)^h \left( \frac{2\pi \zb}{\beta}\right)^{\hb}\mathcal{O}(z,\zb) \,.
\ee
In this paper, we will be interested in computing Lorentzian correlators so the conformal transformation \rref{conftranseucl} becomes
\be \label{conftrans}
z(x,t)=e^{\frac{2\pi}{\beta}(x+t)} \,, \qquad \bar{z}(x,t)=e^{\frac{2\pi}{\beta}(x-t)} \,,
\ee
where $t$ is now Lorentzian time. Note that although $z^*=\zb$ for Euclidean times, it is no longer true in Lorentzian time. However, any Lorentzian correlation function with arbitrary ordering of operators can always be obtained from its Euclidean counterpart upon doing the appropriate analytic continuation. We will describe the procedure shortly. The correlation function we wish to compute is
\be \label{Fdef}
F(t)\equiv\frac{\Braket{V(0,0)W(x,t)V(0,0)W(x,t)}_\beta}{\Braket{V(0,0)V(0,0)}_\beta \Braket{W(x,t)W(x,t)}_\beta} \,,
\ee
which by \rref{conftrans} can be mapped to
\be
F(t)=\frac{\Braket{W(z_1,\zb_1)W(z_2,\zb_2)V(z_3,\zb_3)V(z_4,\zb_4)}}{\Braket{W(z_1,\zb_1)W(z_2,\zb_2)} \Braket{V(z_2,\zb_2)V(z_4,\zb_4)}} \,.
\ee
The positions of the operators are
\begin{alignat}{2}
&z_1=e^{\frac{2\pi}{\beta}(x+t+i \epsilon_1)} \qquad &&\zb_1=e^{\frac{2\pi}{\beta}(x-t-i \epsilon_1)} \notag \\
&z_2=e^{\frac{2\pi}{\beta}(x+t+i \epsilon_2)} \qquad &&\zb_2=e^{\frac{2\pi}{\beta}(x-t-i \epsilon_2)} \notag \\
&z_3= e^{\frac{2\pi}{\beta}(i \epsilon_3)} \qquad &&\zb_3=e^{\frac{2\pi}{\beta}(-i \epsilon_3)}  \\
&z_4=e^{\frac{2\pi}{\beta}(i \epsilon_4)} \qquad &&\zb_4=e^{\frac{2\pi}{\beta}(-i \epsilon_4)} \notag \,.
\end{alignat}
The various factors of $\epsilon_i$ are regulators that are needed to analytically continue the Euclidean correlator to Lorentzian time. The procedure is as follows. We start with finite values of $\epsilon_i$ at $t=0$. This is a Euclidean correlator. We then analytically continue in Lorentzian time by increasing $t$ keeping the $\epsilon_i$ finite. Finally, one can smear the operators over Lorentzian time and then take the $\epsilon_i\to0$. The order in which we take the $\epsilon_i$ to zero will determine the ordering in Lorentzian time.
Similarly to \cite{Roberts:2014ifa}, we will omit this final step and keep the $\epsilon_i$ finite and refer the reader to section 2.4 of \cite{Roberts:2014ifa} for a more detailed discussion.

Note that by conformal symmetry, this function is only a function of the cross ratios given by
\be
u=\frac{z_{12}z_{34}}{z_{13}z_{24}} \,, \qquad \bar{u}=\frac{\zb_{12}\zb_{34}}{\zb_{13}\zb_{24}} \,,
\ee
which means
\be
F(t)=F(u,\bar{u}) \,.
\ee
If the correlator is purely Euclidean, then the function $F(u,\bar{u})$ is single valued. However, this is no longer true for Lorentzian times and $F(u,\bar{u})$ becomes multivalued: a branch cut stretches from 1 to $\infty$. One must specify which sheet we do the computation on, which again is related to the choice of ordering for the operators. The ordering given in \rref{Fdef} corresponds to doing the analytic continuation
\be
1-u\to 1-u \,, \qquad 1-\bar{u}\to e^{2\pi i }(1-\bar{u}) \,,
\ee
which circles around the  branch point $\bar{u}=1$ and hence crosses the branch cut. However, nothing happens for the holomorphic cross-ratio $u$ because we have broken the symmetry between $u$ and $\bar{u}$ by considering Lorentzian times. Note that at late times, the cross-ratios are small and read
\be
u\approx -e^{-\frac{2\pi}{\beta} (t+x)} \epsilon_{12}^* \epsilon_{34} \,, \qquad \bar{u}\approx -e^{-\frac{2\pi}{\beta} (t-x)} \epsilon_{12}^* \epsilon_{34} \,,
\ee
with
\be
\epsilon_{ij}=i \left(e^{\frac{2\pi i}{\beta}\epsilon_i}-e^{\frac{2\pi i}{\beta}\epsilon_j}\right) \,.
\ee

We will now turn to the computation of these correlation function for permutation orbifolds.

\section{Permutation Orbifolds}
We will consider OTO correlation functions in a particular family of 2d CFTs: permutation orbifolds. Permutation orbifolds give a huge landscape of 2d CFTs at large central. They are built in the following way: consider any 2d CFT $\mathcal{C}$ with central charge $c$; we will call $\mathcal{C}$ the \textit{seed} theory. Now consider the $N$-fold tensor product
\be \label{producttheory}
\mathcal{C}^{\otimes N} \,,
\ee
which has central charge $Nc$. This theory has a global $S_N$ symmetry that permutes any of the $N$ copies of $\mathcal{C}$. We may then take a quotient of this product theory by any subgroup of the permutation group $G_N \subseteq S_N$. We are thus led to define
\be
\mathcal{C}_N=\frac{\mathcal{C}^{\otimes N}}{G_N} \,.
\ee
One can define such a theory for any seed theory $\mathcal{C}$ and for any group $G_N$, which by taking $N$ large, gives a huge landscape of 2d CFTs with a large central charge and thus a possible semi-classical holographic dual.

It is important to note that not all permutation groups have a well-defined large $N$ limit. For example, the number of states at fixed dimension $\Delta$ may diverge as $N\to\infty$. We will therefore work with a subset of permutation orbifolds, those for which the group $G_N$ is oligomorphic \cite{MR0401885,MR2581750,MR1066691}. Oligomorphic means that the group has a finite number of orbits on $K$-tuples as $N\to\infty$ which in turn gives a finite number of states \cite{Haehl:2014yla,Belin:2014fna,Belin:2015hwa} of fixed dimensions $\Delta$ as $N\to\infty$. For example, this excludes the cyclic orbifolds $\mathbb{Z}_N$ but allows group quite smaller than $S_N$ such as the wreath product $S_{\sqrt{N}}\wr S_{\sqrt{N}}$.

\subsection{Large N factorization}
In \cite{Belin:2015hwa}, it was shown that a class of permutation orbifolds satisfy large $N$ factorization. This was shown to be the case for the symmetric group, the wreath product, as well as any democratic group, $\ie$ groups with orbits of the same size in the large $N$ limit. In this paper, we will assume that large $N$ factorization holds for all oligomorphic permutation groups. Even if a general proof is still missing, we believe this to be true. As evidence, note that even the cyclic group that is not oligomorphic still satisfies large $N$ factorization. Alternatively, one can consider our results to apply to democratic permutation groups. For simplicity, we will derive the expressions for the symmetric group $S_N$ and only reintroduce the factors counting the numbers of orbits when we discuss the OTO correlators.

We now review the derivation of large $N$ factorization given in \cite{Belin:2015hwa} and generalize it to four point functions. For simplicity, we will only consider untwisted sector operators but the generalization to arbitrary twisted sectors follows trivially. The untwisted sector operators can be described in the following way. Consider an ordered $K$-tuple $\vec{K}$ of
distinct integers, and a $K$-vector $\vec{\varphi}$
of states in the seed theory,
\be
\phi = \phi_{(\vec{K},\vec{\varphi})}\ .
\ee
The notation is that the CFT $K_i$ is in state $\varphi_i$ while all other CFTs are in the vacuum. This is a state of the product theory \rref{producttheory}. To obtain a state invariant under the action of $G_N$, we must sum over images of the group. This gives
\be
\Phi = \sum_{g\in G_N} \phi_{(g.\vec{K},\vec{\varphi})} \,,
\ee
where $g$ acts only on the vector of integers, namely it permutes which of the $N$ copies are in excited states. Any untwisted sector state of the orbifold theory can be expressed this way. Note that states where a single copy is in a non-trivial state correspond to single-trace operators, whereas those that have multiple excited states give multi-trace operators.

We will now prove that a general 4-point function will have the following schematic structure
\be
\frac{\Braket{\Phi_1\Phi_2\Phi_3\Phi_4}}{\mathcal{N}_1\mathcal{N}_2\mathcal{N}_3\mathcal{N}_4}\approx\sum\prod\Braket{\varphi \varphi} +\sum N^{-n_3/2-n_4} \Braket{\varphi \varphi\varphi}^{n_3}\Braket{\varphi \varphi \varphi \varphi}^{n_4}\prod \Braket{\varphi \varphi} \,,\label{factorization}
\ee
where the $\mathcal{N}_i$ are normalization factors and $\braket{\varphi_1...\varphi_k}$ corresponds to a $k$-point function of the seed theory. The leading term in the $1/N$ expansion corresponds to the disconnected contribution for single-trace operators. For multi-trace operators, it simply corresponds to the sum over all Wick contractions \cite{Belin:2017nze}.

It is usefull to consider the following schematic contraction of the different excited factors of the seed theory.

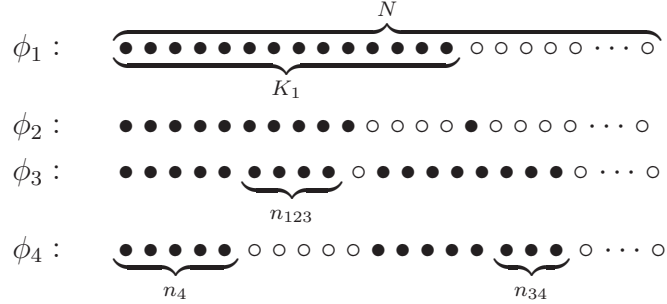
\begin{eqnarray*}
\phi_1:&&\overbrace{\underbrace{\bb\bb\bb\bb\bb\bb\bb\bb\bb\bb\bb\bb\bb\bb}_{K_1}\wb\wb\wb\wb\wb\cdots\wb}^N\\
\phi_2:&&\bb\bb\bb\bb\bb\bb\bb\bb\bb\bb\wb\wb\wb\wb\bb\wb\wb\wb\wb\cdots\wb\\
\phi_3:&&\bb\bb\bb\bb\bb\underbrace{\bb\bb\bb\bb}_{n_{123}}\wb\bb\bb\bb\bb\bb\bb\bb\bb\wb\cdots\wb
\\
\phi_4:&&\underbrace{\bb\bb\bb\bb\bb}_{n_4}\wb\wb\wb\wb\wb\bb\bb\bb\bb\bb\underbrace{\bb\bb\bb}_{n_{34}}\wb\cdots\wb
\end{eqnarray*}
Each black dot corresponds to an excited states whereas white dots correspond to vacua\footnote{In principal, one would have to keep track of the different seed theory states. This would amount to giving different colors to the black dots. This will only keep track of $N$-independent numbers so we neglect it for simplicity. See \cite{Belin:2017nze} for an exact expression in the case of the symmetric group.}. The numbers $n_{ij}$ or $n_{ijk}$ tell us the number of 2 or 3 point overlaps of the seed theory and $n_4$ gives the number of 4 point overlaps. The numbers are not all independent, we have
\bea \label{length}
n_{12}+n_{13}+n_{14}+n_{123}+n_{124}+n_{134}+n_4&=&K_1 \notag \\
n_{12}+n_{23}+n_{24}+n_{123}+n_{124}+n_{234}+n_4&=&K_2 \notag \\
n_{13}+n_{23}+n_{34}+n_{123}+n_{134}+n_{234}+n_4&=&K_3 \\
n_{14}+n_{24}+n_{34}+n_{124}+n_{134}+n_{234}+n_4&=&K_4 \notag \,.
\eea
Each state in this pictorial representation is accompanied by its own sum over the permutation group. To calculate the correlation function, we only need to keep track of the non-zero contributions, which means we only keep the terms where there are at least two states overlapping because any 1-point function would vanish. We will take the states of the seed to be orthonormal such  that
\be
\Braket{\varphi_i\varphi_j}=\delta_{ij} \,.
\ee

We must also take into account the normalization factors $\mathcal{N}_i$ for the four operators $\Phi_i$. The two point function can be shown to be
\be
\Braket{\Phi_i \Phi_i}= N! (N-K_i)! \,,
\ee
which gives
\be
\mathcal{N}_i = \sqrt{N! (N-K_i)!} \,.
\ee
We are now ready to evaluate the contribution to the 4p-function. Keeping only the $N$-dependent factors, we obtain the following contributions:
\begin{itemize}
\item The sum over the group $S_N^1$ simply gives $N!$
\item For $S_N^2$, there are two contributions. First, the excited states of $\phi_2$ that are not contracted with $\phi_1$ can be distributed in any way on the vacua of $\phi_1$. This gives a contribution of $\binom{N-K_1}{K_2-n_{12}-n_{123}-n_{124}-n_4}$. Then, there is a factor of $(N-K_2)!$ coming from the permutation of the vacua of $\phi_2$.
\item For $S_N^3$ There are also two contributions. First the contractions of 3 and 4 only can be distributed in any way along the vacua not occupied by 1 or 2. This gives a contribution of $\binom{N-K_1-n_{23}-n_{24}-n_{234}}{n_{34}}$. The vacua of $\phi_2$ also give a contribution of $(N-K_3)!$.
\item Finally, the only contribution from $S_N^4$ comes from the vacua as all the other contractions are fixed. This gives $(N-K_4)!$
\end{itemize}

Adding the normalization factors, we get a contribution of
\be
\frac{\sqrt{(N-K_1)!(N-K_2)!(N-K_3)!(N-K_4)!}}{N!(N-1/2(K_1+K_2+K_3+K_4)!+n_4+n_3/2)} \,,
\ee
with $n_3=n_{123}+n_{124}+n_{134}+n_{234}$. Using Stirling's approximation, it is easy to see that we recover the form \rref{factorization}. A term which has $n_3$ and $n_4$ three and four-point overlap and will be of order 
\be
N^{-n_3/2+n_4}
\ee
We will now proceed to the evaluation of $F(t)$ using the results we just derived.

\section{Out-of-time-ordered Correlators in Permutation Orbifolds}

\subsection{Untwisted Sector Single-trace Operators}

We will start with the simplest possible choice of operators: single-trace operators. These will be operators given by
\be
V=\sum_{k=1}^N \varphi_i^k \,, \qquad W=\sum_{k=1}^N\varphi_j^k \,,
\ee
where $i$ and $j$ label operators of the seed theory and the sum over $k$ sums over the $N$ copies. They are symmetric operators invariant under $S_N$. 

One might wonder wether there are multiple operators of this form for a given $\phi_i$ if the group is a subgroup of $S_N$, rather than the full symmetric group. This would mean that there are multiple orbits of the group when acting on 1-tuples. While there clearly are examples of oligomorphic permutation groups that have this property (for example $S_{N/2}\times S_{N/2}$), we will not consider them here. These theories would have more than one stress tensor and  would hence be peculiar. We will focus on oligomorphic permutation groups who have a single orbit when acting on 1-tuples.

For such a choice of operators, it is easy to see that there cannot be 3-point overlaps hence $n_3=0$. We get
\be
F(t)=1+\frac{1}{N}\frac{\Braket{\varphi_i(z_1,\zb_1)\varphi_i(z_2,\zb_2)\varphi_j(z_3,\zb_3)\varphi_j(z_4,\zb_4)}}{\Braket{\varphi_i(z_1,\zb_1)\varphi_i(z_2,\zb_2)} \Braket{\varphi_j(z_2,\zb_2)\varphi_j(z_4,\zb_4)}} \,.
\ee

This shows that the dynamics of the four-point function at late times is completely fixed by the behaviour of the OTO correlation function in the seed theory. It is now easy to see that $F(t)$ cannot become small at late times. To see that, notice that there are essentially two scenarios. First, the seed theory could be chaotic. In that case, its own OTO correlator would vanish at late times. Second, it could be non-chaotic which means its OTO would not vanish at late times and stay of order one. In any event, the OTO correlator of the seed theory can never become $\mathcal{O}(N)$. In fact, it cannot even know about the existence of a parameter $N$. This shows that the OTO of all single-trace operators stays of $\mathcal{O}(1)$ at late times. We now turn to multi-trace operators.

\subsection{Untwisted Sector Multi-trace Operators}

For a multi-trace operator, the expression of $F(t)$ will be more complicated. In general it will be of the form
\be \label{Fmultitrace}
F(t)=1+\sum g_2(\vec{K},\vec{\varphi}) f_2(t)+\frac{1}{N}\left[\sum g_3(\vec{K},\vec{\varphi})f_3(t)+\sum g_4(\vec{K},\vec{\varphi})f_4(t)\right]+...
\ee
where $f_2(t),f_3(t),f_4(t)$ are the contribution coming from 2-point, 3-point and 4-point overlaps respectively. The $g_i$ are combinatorial factors that depend on the number of seed operators chosen, wether they are all the same or not, and the choice of the group. It can be determined purely from group theory arguments by counting the number of orbits on a given (un)ordered $K$-tuple. In any case, note that
\be
g_i\sim\mathcal{O}(1) \,,
\ee
as implied by the fact that we are considering oligomorphic permutation groups which by definition have a finite number of orbits as $N\to\infty$. Also note that the sums in \rref{Fmultitrace} run over an $\mathcal{O}(1)$ number of possibilities. This results from the fact that we considered $V$ and $W$ to be "simple" operators, made out of an $\mathcal{O}(1)$ number of seed theory operators. In particular, $V$ and $W$ must have $\Delta\ll N $. We will now analyze the various contributions to \rref{Fmultitrace}.

\subsubsection{2-point Overlaps}

The 2-point overlap captures the disconnected contribution to the four-point function. For single-trace operators, this term simply gave one. For multi-trace operators, there can be contractions between seed operators in $V$ and seed operators in $W$ if $V$ and $W$ share a same seed theory operator. This means we can have terms of the form
\bea \label{f2untw}
\frac{\Braket{\varphi_a (z_1) \varphi_a(z_4)}\Braket{\varphi_a(z_2) \varphi_a(z_3)}}{\Braket{\varphi_a (z_1) \varphi_a(z_2)}\Braket{\varphi_a(z_3) \varphi_a(z_4)}}&=&\left(\frac{u}{1-u}\right)^{2h_a}\left(\frac{\bar{u}}{1-\bar{u}}\right)^{2\bar{h}_a} \notag \\
\frac{\Braket{\varphi_a (z_1) \varphi_a(z_3)}\Braket{\varphi_a(z_2) \varphi_a(z_4)}}{\Braket{\varphi_a (z_1) \varphi_a(z_2)}\Braket{\varphi_a(z_3) \varphi_a(z_4)}}&=&u^{2h_a}\bar{u}^{2\bar{h}_a} \,.
\eea
One can quickly see that these terms do not do anything interesting upon taking an analytic continuation. Terms of the first type simply gives a phase upon analytic continuation whereas those of the second type do nothing. Furthermore, $u,\bar{u}\to0$ when $t\gg t_d$ which means both types of terms will quickly decay. We now turn to the three-point overlaps.

\subsubsection{3-point Overlaps}
First, it is important to note that there must be an even number of three point overlaps because the four point function we consider has two pairs of identical operators. It is then easy to see that the most general form of $f_3(t)$ will be
\be \label{f3untw}
f_3(t)= C_i^2 u^{p_1(h_i)} (1-u)^{p_2(h_i)} \bar{u}^{\bar{p}_1(\bar{h}_i)} (1-\bar{u})^{\bar{p}_2(\bar{h}_i)}
\ee
where the $p_i$ are linear functions of the conformal weights and $C_i^2$ is an ope coefficient of three operators of the seed theory squared. Also, one can easily check that $p_1,\bar{p}_1>0$. This shows that the three-point overlaps do not have interesting analytic continuations and just like the two-point overlaps, they would pick up a simple phase and anyway decay for times much greater than the dissipation time.

\subsubsection{4-point Overlaps}
The contributions from the 4-point overlap are of course very similar to the case where we had single-trace operators. They take the form
\be
\frac{\Braket{\varphi_a (z_1)\varphi_a(z_2) \varphi_b(z_3) \varphi_b(z_4)}}{\Braket{\varphi_a (z_1) \varphi_a(z_2)}\Braket{\varphi_b(z_3) \varphi_b(z_4)}} \,,
\ee
which again is an OTO correlation function in the seed theory. The only difference with the single-trace operators is that there will be a sum over different OTO correlators in the seed theory, weighted by sum combinatorial factor that depends on the number of different states and the group $G_N$.

Nevertheless, this implies that the OTO correlators of the multi-trace operators cannot decrease at late times as it is built out of an $\mathcal{O}(1)$ number of seed theory OTO correlators that each may decay or not, but at least can never become large. This closes our analysis of the untwisted sector operators.

We have shown that an arbitrary choice of untwisted sector operators $V$ and $W$ with $\Delta\ll N$ yields an OTO correlator that cannot decay at late times. It is tempting to conclude that this already proves that permutation orbifolds cannot be chaotic. Note however that the growth of operators in the symmetric product theory is dominated by twisted sector operators \cite{Belin:2014fna}, which grow as
\be
\rho_{\text{tw}}(\Delta)\approx e^{2\pi \Delta} \,,
\ee
whereas the growth of untwisted sector operators only goes as
\be
\rho_{\text{untw}}(\Delta)\approx e^{\frac{\Delta}{\log\Delta}} \,.
\ee
One could then argue that an untwisted sector operator is actually not generic, and that it is perhaps the reason why they do not decay at late times. We will now show that the twisted sector operators behave exactly in the same way.

\subsection{Twisted Sector Operators}

The expression for generic twisted sector operators will still take the form \rref{Fmultitrace}. As showed in the previous section, there is no fundamental difference between single-trace and multi-trace operators at this level so we will consider multi-trace operators to stay as general as possible. It is easy to show that the 2-point and 3-point overlaps behave exactly the same way as for the untwisted sector operators. This results from the fact that the only data necessary do derive \rref{f2untw} and \rref{f3untw} was the conformal weights. $f_3(t)$ also carries an OPE coefficient of the seed theory squared but these are $\mathcal{O}(1)$ numbers and do not play an important role. For this reason, no interesting contribution can come from the two and three-point overlaps.

We still need to consider the four-point overlaps. The general structure of these contributions is sketched in \cite{Belin:2015hwa,Lunin:2000yv,Pakman:2009zz,Burrington:2012yn} and reads
\be \label{OTOtwisted}
\frac{\Braket{\varphi_a (z_1)\varphi_a(z_2) \varphi_b(z_3) \varphi_b(z_4)}}{\Braket{\varphi_a (z_1) \varphi_a(z_2)}\Braket{\varphi_b(z_3) \varphi_b(z_4)}} \,,
\ee
where the $\varphi_{a,b}$ are now operators in some $S_k$ orbifold theory where $k\sim\mathcal{O}(1)$. For example, if the twisted sector contains a single cycle of length $k$, the operators will be twisted sector operators in a $\mathbb{Z}_k$ orbifold theory, which were considered in \cite{Caputa:2017rkm}\footnote{It is also possible to view these twisted sector four-point functions as correlation functions of the seed theory but on a complicated Riemann surface, although it will not be particularly helpful here.}.

But now, the same logic we applied for the untwisted sector operators can be used here. Independently of wether this $S_k$ orbifold theory is chaotic or not, its OTO four-point function can never become $\mathcal{O}(N)$. It would be nice to be able to bound the OTO four-point function, for example by its value at $t=0$. This can be done for the spectral-form factor which is an analytic continuation of the partition function \cite{Dyer:2016pou,Cotler:2016fpe}. The bound simply comes from the fact that the partition function is a sum over positive contributions and introducing phases can only decrease its value. Unfortunately, a similar reasoning does not apply to the OTO four-point function as the Euclidean correlator is not a sum over positive contributions. Also, if $V$ and $W$ are null separated then the Lorentzian correlator must diverge. Nevertheless, once we move away from the lightcones and go to late times the correlator \rref{OTOtwisted} still cannot be of order $N$.

As an example, one can consider the four-point function of two twisted and two untwisted operators in the D1D5 CFT. The two twist operators we will consider are special in that they correspond to the operators that deform the theory towards the strongly coupled regime. The Euclidean correlator was calculated in \cite{Burrington:2012yn} and it was shown in \cite{Perlmutter:2016pkf} that it does not vanish at late times after we analytically continue to the OTO setup. A general four-point function of twisted operators in the D1D5 CFT at the orbifold point will be hard to compute but for the reasons mentioned above, we do not expect it to decay at late times.

This closes our discussion of all possible operators of dimension $\Delta\ll N$ in permutation orbifolds. We have shown that their OTO correlators do not decay at late times, indicating that permutation orbifolds are not chaotic theories. This is in fact expected, since permutation orbifolds correspond to free discrete gauge theories.

\subsection{The Early Time Behaviour}
So far, we have been interested in the behaviour of the OTO correlator at late times ($t\sim t_s$) and have simply investigated wether it decays to zero or not. In this sense, permutation orbifolds are universal and all share the same structure: the OTO of generic operators does not decay at late times and stays of order one.

However, one may wonder what happens at earlier times ($t\ll t_s$). This is where the universality will break down and the physics will be theory-dependent. In particular, the choice of the seed theory will dictate the dynamics at early times. This dependence on details of the theory was made explicit in \cite{Caputa:2017rkm} where the authors considered the behaviour of the OTO correlator in $\mathbb{Z}_n$ orbifolds of $\mathbb{T}^2$. As they show, the answer depends strongly on the compactification radius and wether it is a rational number or not. For irrational compactification radii, they find a polynomial decay. This is directly relevant for the behaviour of the OTO correlator in permutation orbifolds, where this type of behaviour will be relevant at early times (see also \cite{Caputa:2016tgt} for a discussion of OTO correlators in rational CFTs).

In general, if one picks a seed theory that is chaotic, we expect there to be some interesting time-dependence at early times dictated by the physics of the seed theory. Note however that there is no clear notion of a Lyapunov exponent for CFTs with central charge $c\sim \mathcal{O}(1)$ as there is no parametrically large difference between the dissipation time and the scrambling time.

\subsection{Deformation to Strong Coupling}
We know that CFTs dual to weakly-coupled supergravity should be maximally chaotic since black holes in Einstein Gravity saturate the chaos bound \cite{Maldacena:2015waa}. This means that deforming the D1D5 CFT away from the orbifold point should drastically change the behaviour of the OTO correlation function. To see this, one needs to study the orbifold theory deformed by a twist-2 operator. The deformation is
\be
\delta S=\alpha \int dz d\bar{z} O(z,\bar{z})
\ee
where $O(z,\zb)$ is the exactly marginal operator described in \cite{Burrington:2012yn,Avery:2010er,Gaberdiel:2015uca} built from the twist-2 operator. The scaling of the coupling $\alpha$ can be shown to be \cite{Pakman:2009zz}
\be
\alpha \sim \lambda N^{1/2}
\ee
where $\lambda$ is the 't Hooft coupling and is fixed in the limit $N\to\infty$. Only even powers of $\lambda$ will appear in the OTO correlators as we do conformal perturbation theory. This means that the first correction will be of the form
\be
\lambda^2 N \frac{\braket{VWVW \sigma \sigma}_{\text{con}}}{\braket{VV}\braket{WW}} \,.
\ee
It would be very interesting to compute this correction using second-order conformal perturbation theory. Nonetheless, much more work is needed to compute the Lyapunov exponent perturbatively. One would need to resum ladder diagrams along the lines of \cite{Stanford:2015owe}, which in this case means considering four-point function on arbitrary genus Riemann surfaces, integrated over the moduli of the surface. This appears to be a very complicated task and it is not clear to us how to use the ladder diagram re-organization in this context. It would be interesting to attempt this calculation but we leave this for future work.

\section*{Acknowledgements}
I would like to thank Fotis Dimitrakopoulos, Guy Gur-Ari, Christoph Keller, Dan Roberts, Gabor Sarosi, Steve Shenker and Ida Zadeh for useful discussions. I am especially grateful to Nathan Benjamin and Ethan Dyer for collaboration at an early stage of this project. I thank the Galileo Galilei Institute for Theoretical Physics (GGI) for the hospitality and INFN for partial support during the completion of this  work,  within  the  program \textit{New Developments in AdS$_3$/CFT$_2$ Holography}. I  am supported by the Foundation for Fundamental Research on Matter (FOM). This work is part of the $\Delta$-ITP consortium, a program of the NWO funded by the Dutch Ministry of Education, Culture and Science (OCW).

\bibliographystyle{ytphys}
\bibliography{ref}

\providecommand{\href}[2]{#2}\begingroup\raggedright\begin{thebibliography}{10}

\bibitem{Shenker:2013pqa}
S.~H. Shenker and D.~Stanford, ``{Black holes and the butterfly effect},''
  \href{http://dx.doi.org/10.1007/JHEP03(2014)067}{{\em JHEP} {\bfseries 03}
  (2014) 067},
\href{http://arxiv.org/abs/1306.0622}{{\ttfamily arXiv:1306.0622 [hep-th]}}.

\bibitem{Shenker:2013yza}
S.~H. Shenker and D.~Stanford, ``{Multiple Shocks},''
  \href{http://dx.doi.org/10.1007/JHEP12(2014)046}{{\em JHEP} {\bfseries 12}
  (2014) 046},
\href{http://arxiv.org/abs/1312.3296}{{\ttfamily arXiv:1312.3296 [hep-th]}}.

\bibitem{Leichenauer:2014nxa}
S.~Leichenauer, ``{Disrupting Entanglement of Black Holes},''
  \href{http://dx.doi.org/10.1103/PhysRevD.90.046009}{{\em Phys. Rev.}
  {\bfseries D90} no.~4, (2014) 046009},
\href{http://arxiv.org/abs/1405.7365}{{\ttfamily arXiv:1405.7365 [hep-th]}}.

\bibitem{Kitaevtalks}
A.~Kitaev, ``{A simple model of quantum holography},''.

\bibitem{Roberts:2014isa}
D.~A. Roberts, D.~Stanford, and L.~Susskind, ``{Localized shocks},''
  \href{http://dx.doi.org/10.1007/JHEP03(2015)051}{{\em JHEP} {\bfseries 03}
  (2015) 051},
\href{http://arxiv.org/abs/1409.8180}{{\ttfamily arXiv:1409.8180 [hep-th]}}.

\bibitem{Roberts:2014ifa}
D.~A. Roberts and D.~Stanford, ``{Two-dimensional conformal field theory and
  the butterfly effect},''
  \href{http://dx.doi.org/10.1103/PhysRevLett.115.131603}{{\em Phys. Rev.
  Lett.} {\bfseries 115} no.~13, (2015) 131603},
\href{http://arxiv.org/abs/1412.5123}{{\ttfamily arXiv:1412.5123 [hep-th]}}.

\bibitem{Jackson:2014nla}
S.~Jackson, L.~McGough, and H.~Verlinde, ``{Conformal Bootstrap, Universality
  and Gravitational Scattering},''
  \href{http://dx.doi.org/10.1016/j.nuclphysb.2015.10.013}{{\em Nucl. Phys.}
  {\bfseries B901} (2015) 382--429},
\href{http://arxiv.org/abs/1412.5205}{{\ttfamily arXiv:1412.5205 [hep-th]}}.

\bibitem{Shenker:2014cwa}
S.~H. Shenker and D.~Stanford, ``{Stringy effects in scrambling},''
  \href{http://dx.doi.org/10.1007/JHEP05(2015)132}{{\em JHEP} {\bfseries 05}
  (2015) 132},
\href{http://arxiv.org/abs/1412.6087}{{\ttfamily arXiv:1412.6087 [hep-th]}}.

\bibitem{Maldacena:2015waa}
J.~Maldacena, S.~H. Shenker, and D.~Stanford, ``{A bound on chaos},''
\href{http://arxiv.org/abs/1503.01409}{{\ttfamily arXiv:1503.01409 [hep-th]}}.

\bibitem{Polchinski:2015cea}
J.~Polchinski, ``{Chaos in the black hole S-matrix},''
\href{http://arxiv.org/abs/1505.08108}{{\ttfamily arXiv:1505.08108 [hep-th]}}.

\bibitem{Caputa:2015waa}
``{Scrambling time from local perturbations of the eternal BTZ black hole},''
  \href{http://dx.doi.org/10.1007/JHEP08(2015)011}{{\em JHEP} {\bfseries 08}
  (2015) 011},
\href{http://arxiv.org/abs/1503.08161}{{\ttfamily arXiv:1503.08161 [hep-th]}}.

\bibitem{Hosur:2015ylk}
P.~Hosur, X.-L. Qi, D.~A. Roberts, and B.~Yoshida, ``{Chaos in quantum
  channels},'' \href{http://dx.doi.org/10.1007/JHEP02(2016)004}{{\em JHEP}
  {\bfseries 02} (2016) 004},
\href{http://arxiv.org/abs/1511.04021}{{\ttfamily arXiv:1511.04021 [hep-th]}}.

\bibitem{Stanford:2015owe}
D.~Stanford, ``{Many-body chaos at weak coupling},''
  \href{http://dx.doi.org/10.1007/JHEP10(2016)009}{{\em JHEP} {\bfseries 10}
  (2016) 009},
\href{http://arxiv.org/abs/1512.07687}{{\ttfamily arXiv:1512.07687 [hep-th]}}.

\bibitem{Gur-Ari:2015rcq}
G.~Gur-Ari, M.~Hanada, and S.~H. Shenker, ``{Chaos in Classical D0-Brane
  Mechanics},'' \href{http://dx.doi.org/10.1007/JHEP02(2016)091}{{\em JHEP}
  {\bfseries 02} (2016) 091},
\href{http://arxiv.org/abs/1512.00019}{{\ttfamily arXiv:1512.00019 [hep-th]}}.

\bibitem{Berkowitz:2016znt}
E.~Berkowitz, M.~Hanada, and J.~Maltz, ``{Chaos in Matrix Models and Black Hole
  Evaporation},'' \href{http://dx.doi.org/10.1103/PhysRevD.94.126009}{{\em
  Phys. Rev.} {\bfseries D94} no.~12, (2016) 126009},
\href{http://arxiv.org/abs/1602.01473}{{\ttfamily arXiv:1602.01473 [hep-th]}}.

\bibitem{Polchinski:2016xgd}
J.~Polchinski and V.~Rosenhaus, ``{The Spectrum in the Sachdev-Ye-Kitaev
  Model},'' \href{http://dx.doi.org/10.1007/JHEP04(2016)001}{{\em JHEP}
  {\bfseries 04} (2016) 001},
\href{http://arxiv.org/abs/1601.06768}{{\ttfamily arXiv:1601.06768 [hep-th]}}.

\bibitem{Fitzpatrick:2016thx}
A.~L. Fitzpatrick and J.~Kaplan, ``{A Quantum Correction To Chaos},''
  \href{http://dx.doi.org/10.1007/JHEP05(2016)070}{{\em JHEP} {\bfseries 05}
  (2016) 070},
\href{http://arxiv.org/abs/1601.06164}{{\ttfamily arXiv:1601.06164 [hep-th]}}.

\bibitem{Michel:2016kwn}
B.~Michel, J.~Polchinski, V.~Rosenhaus, and S.~J. Suh, ``{Four-point function
  in the IOP matrix model},''
  \href{http://dx.doi.org/10.1007/JHEP05(2016)048}{{\em JHEP} {\bfseries 05}
  (2016) 048},
\href{http://arxiv.org/abs/1602.06422}{{\ttfamily arXiv:1602.06422 [hep-th]}}.

\bibitem{Perlmutter:2016pkf}
E.~Perlmutter, ``{Bounding the Space of Holographic CFTs with Chaos},''
  \href{http://dx.doi.org/10.1007/JHEP10(2016)069}{{\em JHEP} {\bfseries 10}
  (2016) 069},
\href{http://arxiv.org/abs/1602.08272}{{\ttfamily arXiv:1602.08272 [hep-th]}}.

\bibitem{Turiaci:2016cvo}
G.~Turiaci and H.~Verlinde, ``{On CFT and Quantum Chaos},''
  \href{http://dx.doi.org/10.1007/JHEP12(2016)110}{{\em JHEP} {\bfseries 12}
  (2016) 110},
\href{http://arxiv.org/abs/1603.03020}{{\ttfamily arXiv:1603.03020 [hep-th]}}.

\bibitem{Caputa:2017rkm}
P.~Caputa, Y.~Kusuki, T.~Takayanagi, and K.~Watanabe, ``{Out-of-Time-Ordered
  Correlators in $(T^2)^n/\mathbb{Z}_n$},''
\href{http://arxiv.org/abs/1703.09939}{{\ttfamily arXiv:1703.09939 [hep-th]}}.

\bibitem{Sekino:2008he}
Y.~Sekino and L.~Susskind, ``{Fast Scramblers},''
  \href{http://dx.doi.org/10.1088/1126-6708/2008/10/065}{{\em JHEP} {\bfseries
  10} (2008) 065},
\href{http://arxiv.org/abs/0808.2096}{{\ttfamily arXiv:0808.2096 [hep-th]}}.

\bibitem{Lashkari:2011yi}
N.~Lashkari, D.~Stanford, M.~Hastings, T.~Osborne, and P.~Hayden, ``{Towards
  the Fast Scrambling Conjecture},''
  \href{http://dx.doi.org/10.1007/JHEP04(2013)022}{{\em JHEP} {\bfseries 04}
  (2013) 022},
\href{http://arxiv.org/abs/1111.6580}{{\ttfamily arXiv:1111.6580 [hep-th]}}.

\bibitem{Afkhami-Jeddi:2016ntf}
N.~Afkhami-Jeddi, T.~Hartman, S.~Kundu, and A.~Tajdini, ``{Einstein gravity
  3-point functions from conformal field theory},''
\href{http://arxiv.org/abs/1610.09378}{{\ttfamily arXiv:1610.09378 [hep-th]}}.

\bibitem{Belin:2014fna}
A.~Belin, C.~A. Keller, and A.~Maloney, ``{String Universality for Permutation
  Orbifolds},'' \href{http://dx.doi.org/10.1103/PhysRevD.91.106005}{{\em Phys.
  Rev.} {\bfseries D91} no.~10, (2015) 106005},
\href{http://arxiv.org/abs/1412.7159}{{\ttfamily arXiv:1412.7159 [hep-th]}}.

\bibitem{Haehl:2014yla}
F.~M. Haehl and M.~Rangamani, ``{Permutation orbifolds and holography},''
  \href{http://dx.doi.org/10.1007/JHEP03(2015)163}{{\em JHEP} {\bfseries 03}
  (2015) 163},
\href{http://arxiv.org/abs/1412.2759}{{\ttfamily arXiv:1412.2759 [hep-th]}}.

\bibitem{Belin:2015hwa}
A.~Belin, C.~A. Keller, and A.~Maloney, ``{Permutation Orbifolds in the large N
  Limit},'' \href{http://dx.doi.org/10.1007/s00023-016-0529-y}{{\em Annales
  Henri Poincare} (2016) 1--29},
\href{http://arxiv.org/abs/1509.01256}{{\ttfamily arXiv:1509.01256 [hep-th]}}.

\bibitem{Keller:2011xi}
C.~A. Keller, ``{Phase transitions in symmetric orbifold CFTs and
  universality},'' \href{http://dx.doi.org/10.1007/JHEP03(2011)114}{{\em JHEP}
  {\bfseries 03} (2011) 114},
\href{http://arxiv.org/abs/1101.4937}{{\ttfamily arXiv:1101.4937 [hep-th]}}.

\bibitem{Hartman:2014oaa}
T.~Hartman, C.~A. Keller, and B.~Stoica, ``{Universal Spectrum of 2d Conformal
  Field Theory in the Large c Limit},''
  \href{http://dx.doi.org/10.1007/JHEP09(2014)118}{{\em JHEP} {\bfseries 09}
  (2014) 118},
\href{http://arxiv.org/abs/1405.5137}{{\ttfamily arXiv:1405.5137 [hep-th]}}.

\bibitem{Kraus:2017kyl}
P.~Kraus, A.~Sivaramakrishnan, and R.~Snively, ``{Black holes from CFT:
  Universality of correlators at large c},''
  \href{http://dx.doi.org/10.1007/JHEP08(2017)084}{{\em JHEP} {\bfseries 08}
  (2017) 084},
\href{http://arxiv.org/abs/1706.00771}{{\ttfamily arXiv:1706.00771 [hep-th]}}.

\bibitem{Balasubramanian:2016ids}
V.~Balasubramanian, B.~Craps, B.~Czech, and G.~Sárosi, ``{Echoes of chaos from
  string theory black holes},''
  \href{http://dx.doi.org/10.1007/JHEP03(2017)154}{{\em JHEP} {\bfseries 03}
  (2017) 154},
\href{http://arxiv.org/abs/1612.04334}{{\ttfamily arXiv:1612.04334 [hep-th]}}.

\bibitem{Belin:2016yll}
A.~Belin, J.~de~Boer, J.~Kruthoff, B.~Michel, E.~Shaghoulian, and M.~Shyani,
  ``{Universality of sparse $d > 2$ conformal field theory at large $N$},''
  \href{http://dx.doi.org/10.1007/JHEP03(2017)067}{{\em JHEP} {\bfseries 03}
  (2017) 067},
\href{http://arxiv.org/abs/1610.06186}{{\ttfamily arXiv:1610.06186 [hep-th]}}.

\bibitem{MR0401885}
P.~J. Cameron, ``Transitivity of permutation groups on unordered sets,'' {\em
  Math. Z.} {\bfseries 148} no.~2, (1976) 127--139.

\bibitem{MR2581750}
P.~J. Cameron,
  \href{http://dx.doi.org/10.1142/9789814273657_0003}{``Oligomorphic
  permutation groups,''} in {\em Perspectives in mathematical sciences. {II}},
  vol.~8 of {\em Stat. Sci. Interdiscip. Res.}, pp.~37--61.
\newblock World Sci. Publ., Hackensack, NJ, 2009.
\newblock \url{http://dx.doi.org/10.1142/9789814273657_0003}.

\bibitem{MR1066691}
P.~J. Cameron, \href{http://dx.doi.org/10.1017/CBO9780511549809}{{\em
  Oligomorphic permutation groups}}, vol.~152 of {\em London Mathematical
  Society Lecture Note Series}.
\newblock Cambridge University Press, Cambridge, 1990.
\newblock \url{http://dx.doi.org/10.1017/CBO9780511549809}.

\bibitem{Belin:2017nze}
A.~Belin, C.~A. Keller, and I.~G. Zadeh, ``{Genus Two Partition Functions and
  Renyi Entropies of Large c CFTs},''
\href{http://arxiv.org/abs/1704.08250}{{\ttfamily arXiv:1704.08250 [hep-th]}}.

\bibitem{Lunin:2000yv}
O.~Lunin and S.~D. Mathur, ``{Correlation functions for M**N / S(N)
  orbifolds},'' \href{http://dx.doi.org/10.1007/s002200100431}{{\em Commun.
  Math. Phys.} {\bfseries 219} (2001) 399--442},
\href{http://arxiv.org/abs/hep-th/0006196}{{\ttfamily arXiv:hep-th/0006196
  [hep-th]}}.

\bibitem{Pakman:2009zz}
A.~Pakman, L.~Rastelli, and S.~S. Razamat, ``{Diagrams for Symmetric Product
  Orbifolds},'' \href{http://dx.doi.org/10.1088/1126-6708/2009/10/034}{{\em
  JHEP} {\bfseries 10} (2009) 034},
\href{http://arxiv.org/abs/0905.3448}{{\ttfamily arXiv:0905.3448 [hep-th]}}.

\bibitem{Burrington:2012yn}
B.~A. Burrington, A.~W. Peet, and I.~G. Zadeh, ``{Twist-nontwist correlators in
  $M^N/S_N$ orbifold CFTs},''
  \href{http://dx.doi.org/10.1103/PhysRevD.87.106008}{{\em Phys. Rev.}
  {\bfseries D87} no.~10, (2013) 106008},
\href{http://arxiv.org/abs/1211.6689}{{\ttfamily arXiv:1211.6689 [hep-th]}}.

\bibitem{Dyer:2016pou}
E.~Dyer and G.~Gur-Ari, ``{2D CFT Partition Functions at Late Times},''
\href{http://arxiv.org/abs/1611.04592}{{\ttfamily arXiv:1611.04592 [hep-th]}}.

\bibitem{Cotler:2016fpe}
J.~S. Cotler, G.~Gur-Ari, M.~Hanada, J.~Polchinski, P.~Saad, S.~H. Shenker,
  D.~Stanford, A.~Streicher, and M.~Tezuka, ``{Black Holes and Random
  Matrices},''
\href{http://arxiv.org/abs/1611.04650}{{\ttfamily arXiv:1611.04650 [hep-th]}}.

\bibitem{Caputa:2016tgt}
P.~Caputa, T.~Numasawa, and A.~Veliz-Osorio, ``{Out-of-time-ordered correlators
  and purity in rational conformal field theories},''
  \href{http://dx.doi.org/10.1093/ptep/ptw157}{{\em PTEP} {\bfseries 2016}
  no.~11, (2016) 113B06},
\href{http://arxiv.org/abs/1602.06542}{{\ttfamily arXiv:1602.06542 [hep-th]}}.

\bibitem{Avery:2010er}
S.~G. Avery, B.~D. Chowdhury, and S.~D. Mathur, ``{Deforming the D1D5 CFT away
  from the orbifold point},''
  \href{http://dx.doi.org/10.1007/JHEP06(2010)031}{{\em JHEP} {\bfseries 06}
  (2010) 031},
\href{http://arxiv.org/abs/1002.3132}{{\ttfamily arXiv:1002.3132 [hep-th]}}.

\bibitem{Gaberdiel:2015uca}
M.~R. Gaberdiel, C.~Peng, and I.~G. Zadeh, ``{Higgsing the stringy higher spin
  symmetry},'' \href{http://dx.doi.org/10.1007/JHEP10(2015)101}{{\em JHEP}
  {\bfseries 10} (2015) 101},
\href{http://arxiv.org/abs/1506.02045}{{\ttfamily arXiv:1506.02045 [hep-th]}}.

\end{thebibliography}\endgroup

\end{document}